\documentclass[pre,showpacs,twocolumn,superscriptaddress]{revtex4}
\usepackage{epsfig}
\usepackage{graphicx}
\usepackage{bm}
\usepackage{amsmath,amsfonts,amssymb}

\newcommand{\Xlink}{\text{Xlink}}
\newcommand{\ev}{\text{ev}}
\newcommand{\e}{\mathop{\mathrm{e}}\nolimits}
\newcommand{\dd}{\mathrm{d}}
\newcommand{\HRS}{\text{HRS}}
\renewcommand{\vec}[1]{{\bf #1}}

\begin{document}

\title{Force-extension relation of cross-linked anisotropic polymer networks}

\author{Panayotis Benetatos} \affiliation{Cavendish Laboratory, University of Cambridge, 19 J.~J.~Thomson
  Avenue, Cambridge, CB3 0HE, United Kingdom}

\author{Stephan Ulrich} \affiliation{Lorentz Institute for Theoretical Physics, Leiden, Netherlands} \affiliation{Institute for Theoretical
  Physics, Georg-August-Universit\"at G\"ottingen,
  Friedrich-Hund-Platz 1, 37077 G\"ottingen, Germany}

\author{Annette Zippelius} \affiliation{Institute for Theoretical
  Physics, Georg-August-Universit\"at G\"ottingen,
  Friedrich-Hund-Platz 1, 37077 G\"ottingen, Germany} \affiliation{Max
  Planck Institute for Dynamics \& Self-Organization, Bunsenstra{\ss}e
  10, 37073 G\"ottingen, Germany}

\date{\today}

\begin{abstract}
Cross-linked polymer networks with orientational order constitute a
wide class of soft materials and are relevant to biological systems
(e.g., F-actin bundles). We analytically study the nonlinear
force-extension relation of an array of parallel-aligned, strongly
stretched semiflexible polymers with random cross-links. In the
strong stretching limit, the effect of the cross-links is purely
entropic, independent of the bending rigidity of the
chains. Cross-links enhance the differential stretching stiffness of
the bundle. For hard cross-links, the cross-link contribution to the
force-extension relation scales inversely proportional to the
force. Its dependence on the cross-link density, close to the gelation
transition, is the same as that of the shear modulus. The qualitative
behavior is captured by a toy model of two chains with a single
cross-link in the middle.

\end{abstract}
\pacs{82.70.Gg, 64.70.Md, 61.43.Fs, 87.16.Ka}
\maketitle

\section{Introduction}

Anisotropic networks are abundant in nature and among man-made
materials. For example, nematic elastomers \cite{WT} or actin-myosin
assemblies in cells \cite{Cooper}. These networks are constructed from
anisotropic building blocks, nematogens such as stiff rods or
semiflexible polymers which are well modelled by wormlike chains
(WLCs) \cite{STY}. Isotropic networks of WLCs have been studied
extensively in experiment \cite{Kroy_Bausch}, driven by the interest
in biological networks such as the cytoskeleton or the extracellular
matrix. Theoretical approaches have dealt with entangled solution
\cite{Morse} as well as chemically cross-linked networks of WLCs
\cite{Head,Wilhelm}. The former build on the tube model whereas the
latter generalise concepts from rubber elasticity. Aligning linkers
can give rise to a variety of morphologies which have been studied by
means of a generalised Onsager approach \cite{Borukhov} as well as
within a microscopic model \cite{BZ,Kiemes}. Of particular interest
are bundles of filaments which occur in a broad range of cytosceletal
structures and show much richer elastic behaviour than the usual WLC
\cite{Claessens,Lieleg,Heussinger}. Two parallel-aligned, stretched
semiflexible filaments cross-linked by a motor cluster have been used
as the minimal elastic element of an active gel \cite{MCM}.

\begin{figure}
\includegraphics[width=0.5\textwidth]{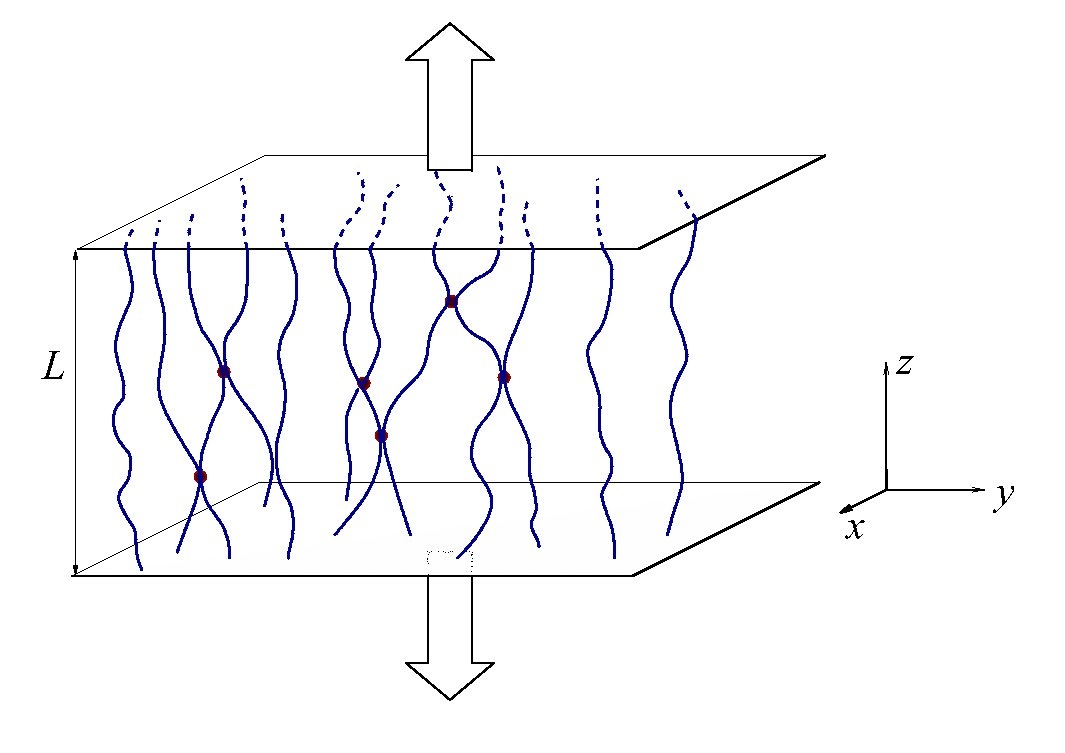} 
\caption{network of aligned wormlike chains under pulling force}
\label{fig:network}
\end{figure}

Here, we consider an anisotropic network of WLCs which have been
aligned along a preferred axis chosen as the $z$--direction. The
alignment is not due to cross-links which we model as springs. Instead
possible mechanisms for alignment are a nematic environment, pulling
forces, grafting surfaces or the Onsager mechanism. A sketch of such a
network is shown in Fig.~\ref{fig:network}. If the alignment is
strong, the WLC model can be replaced by a weakly bending chain (WBC)
as first suggested by Marco and Siggia \cite{MS} for DNA
molecules. The advantage for an analytical approach is enormous
because the single chain model is Gaussian. Our focus in this paper
lies on the force-extension curve of a randomly cross-linked
anisotropic network of strongly aligned filaments. We first consider a
toy model, consisting of two cross-linked filaments. The model allows
us to disentangle the contributions to the effective extension which
are due to either bending stiffnes or cross-links. Subsequently, we
analyse the effects of cross-links for a macroscopic network.

\section{Model}

Our starting point is the energy of a stretched WLC in terms of the
tangent vector ${\bf t}(s)=\frac{d {\bf r}(s)}{ds}=
(\frac{d {\bf r}_{\perp}}{ds},\frac{dz}{ds})\;$,
\begin{equation}
{\cal H}[{\bf
    t} (s) ]= \frac{\kappa}{2} \int_0^L ds \Big(\frac{d {\bf t}
    (s)}{ds}\Big)^2-f\int_0^L ds\frac{dz}{ds}\;.
\end{equation}
Here $\kappa$ denotes the bending stiffnes which is related to the
persistence length $L_p$ via $L_p= 2\kappa/((d-1)k_B T)$, where $d$ is
the dimensionality of the embedding space.  The pulling force is
denoted by $f$, and $0\leq s \leq L$ is the arclength. The local
inextensibility constraint of the WLC is expressed by the condition
$|{\bf t}(s)|=1$. We assume that the chain is strongly stretched so
that tilting of the tangent vector away from the $z$--axis is small
and we can use the approximation
\begin{equation}
\frac{dz}{ds}=\sqrt{1- \Big(\frac{d {\bf r}_{\perp}}{ds} \Big)^2}
\approx 1-\frac{1}{2} \Big(\frac{d{\bf r}_{\perp}}{ds} \Big)^2
\end{equation}
leading to the weakly bending model introduced by Marco and Siggia \cite{MS}
\begin{equation}\label{WBCH}
{\cal H}_0[{\bf
    r}_{\perp} (s) ]= \frac{\kappa}{2} \int_0^L ds 
\Big(\frac{d^2 {\bf r}_{\perp}
    (s)}{ds^2}\Big)^2+\frac{f}{2}\int_0^L ds\Big(\frac{d {\bf r}_{\perp}
    (s)}{ds}\Big)^2
\end{equation}
The central quantity of interest is the extension of the chain
under an applied force $f$, which in the weakly bending approximation
is computed from the thermal fluctuations transverse to the aligning direction:
\begin{eqnarray}
\langle z(L)-z(0) \rangle &=& \int_0^L ds {\Big \langle} \frac{dz}{ds}
{\Big \rangle}\nonumber\\ 
&=& \int_0^L ds {\Big (}1- {\Big \langle} \frac{1}{2}
{\Big (}\frac{d{\bf r}_{\perp}}{ds}{\Big )}^2 {\Big \rangle}{\Big)}
\end{eqnarray}


\section{Toy model: 2 cross-linked chains}

\begin{figure}
  \includegraphics[width=0.3\textwidth]{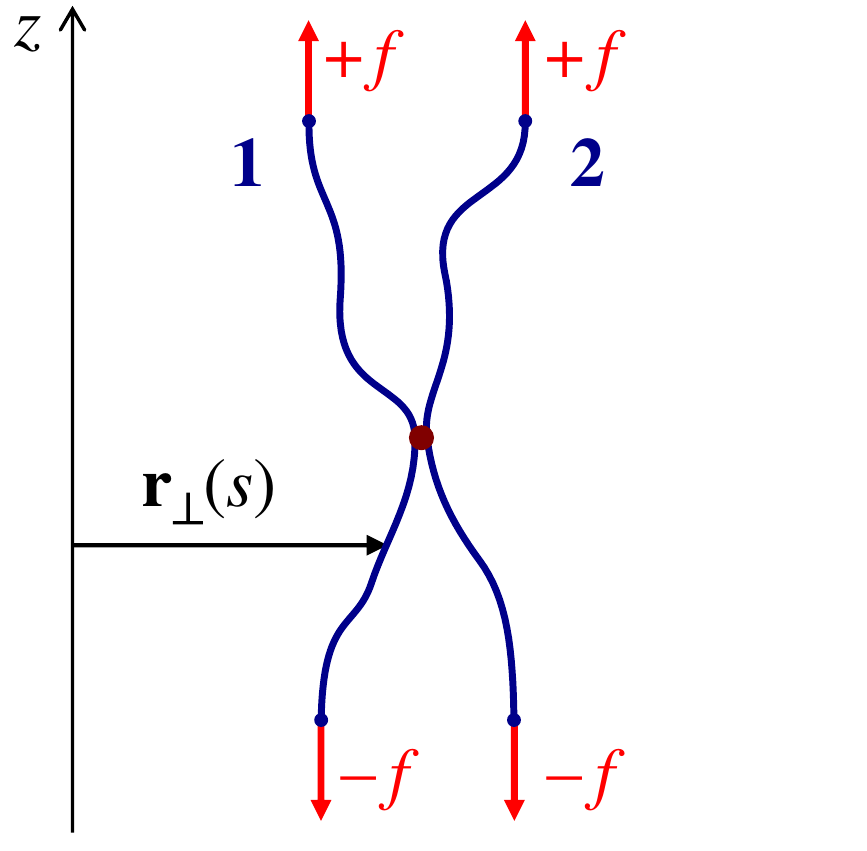}
  \caption{Two aligned chains with one cross-link in the
    middle}\label{fig:two_chains}
\end{figure}

Before addressing the full problem of a randomly cross-linked array of
aligned chains, we discuss the much simpler case of two strongly stretched
chains in two dimensions with one cross-link in the middle, see
Fig.~\ref{fig:two_chains}.
In two dimensions, ${\bf r}_{\perp}(s)=y(s)$ and the Hamiltonian reads
\begin{equation}
{\cal H}={\cal H}_0[y_1(s)]+{\cal
  H}_0[y_1(s)]+\frac{g}{2}\Big(y_1({L}/{2})-y_2({L}/{2})\Big)^2
\end{equation}
The cross-link is modeled as a harmonic spring of stiffness $g$. For
simplicity we impose hinged-hinged boundary conditions:
$y_1(0)=y_1(L)=0,\; y_2(0)=y_2(L)=D$ and $y''_1(0)=y''_1(L)=0,\;
y''_2(0)=y''_2 (L)=0$ for the two chains which are a distance $D$
apart (the prime denotes derivative with respect to $s$).
According to the boundary conditions that we use, the eigenfunction
representation should be
\begin{align*}
y_1(s)=\sum_{l=1}^{\infty}A_l \sin(q_l s)\\
y_2(s)-D=\sum_{l=1}^{\infty}B_l \sin(q_l s)
\end{align*}
and wavenumbers are restricted to values $q_l=\frac{\pi}{L}l\;$, $l \in \mathbb{Z}$. These eigenfunctions diagonalise
the Hamiltonian of the weakly bending chain, whereas the cross-link
gives rise to a term which is quadratic in the amplitudes but not
diagonal. 
Introducing vectors 
\begin{eqnarray*} 
\Gamma&=&(A_1,B_1,A_2,...)\;,\\
u&=&g^{1/2}(\sin(q_1 L/2),\;-\sin(q_1
L/2),\; \sin(q_2 L/2),...)
\end{eqnarray*}
and matrices
\begin{gather*}
C=\frac{L}{2}
\begin{pmatrix}
q_1^2(\kappa q_1^2+f) & 0 & 0 & \hdots \\
0 & q_1^2(\kappa q_1^2+f) & 0 & \hdots \\
0 & 0 & q_2^2(\kappa q_1^2+f) &  \hdots \\
\vdots & \vdots & \vdots & \ddots
\end{pmatrix}
\end{gather*}
the Hamiltonian is rewritten as
\begin{equation}
  H=\frac{1}{2}\sum_{l,m=1}^{\infty}\Gamma_l G_{l,m} \Gamma_m +D g^{1/2}\sum_{l=1}^{\infty} u_l \Gamma_l
\end{equation}
with $G_{l,m}=C_{l,m}+u_l u_m.$ The matrix $G$ is easily inverted:
\begin{equation}
G^{-1}=C^{-1}-\frac{C^{-1}u u^{T}C^{-1}}{1+ u^{T}C^{-1} u}\;, 
\end{equation}
so that we can compute the force-extension curve, say of chain 1,
exactly
\begin{equation}
\langle z_1(L) \rangle
\;=\;L-\frac{L}{4}\sum_l q_l^2 \langle A_l^2 \rangle\;.
\end{equation}

Our model has three characteristic energies: the bending energy,
$\kappa/L$, the work done by the external force, $fL$ and the thermal
energy, $T$ (we set $k_B\equiv 1$). Use of the weakly bending
approximation requires $\kappa/L \gg T$ (which is equivalent to $L_p/L\gg
1$) or $fL\gg T$. This leaves us with one free parameter,
$x:=fL/(\kappa/L)$, namely the ratio of work done by the external
force to bending energy.
This dimensionless quantity can also be interpreted as the squared
ratio of two lengthscales: the total contour length $L$ to the length
$\sqrt{\kappa /f}$ over which the boundary conditions penetrate into
the bulk (i.e., the size of a link in an effective freely-jointed
chain) \cite{PBET}. Actually, we have two additional lengthscales, the
distance between chains, $D$, which we assume to be negligible and the
length of the cross-link ($\propto 1/{\sqrt g})$ or alternatively the
energy of a cross-link relative to the work done by the pulling force,
$gL/f$. 

The result for general $x$ and cross-link strength $g$ 
\begin{equation}
\label{full}
\frac{\langle z_1 \rangle}{L}-1\;=\;-\Delta_{MS}-\Delta_{XL}
\end{equation}
can be decomposed into a contribution, $\Delta_{MS}$, which is characteristic for
a weakly bending chain and well-known from the work of Marko and
Siggia \cite{MS} and a contribution due to the cross-link,
$\Delta_{XL}$
\begin{eqnarray*}
\label{all_forces}
\Delta_{XL}&=&-\frac{gT}{16f^2}\frac{u_2(x)}{u_0(x)}\\
  u_0(x)&=&\sqrt{x}+\frac{gL}{2f}\Big(\sqrt{x}-2\tanh(\sqrt{x}/2)\Big)\\
  u_2(x)&=& 2\sqrt{x}+\sqrt{x}/(\cosh({\sqrt{x}/2}))^2
    -6\tanh(\sqrt{x}/2)
\end{eqnarray*}
In the limit of large $x$, the cross-link contribution is given by
\begin{equation}
\Delta_{XL}=-\frac{gT}{8f^2} \left(1+\frac{gL}{2f} \right)^{-1}.
\end{equation}
For soft cross-links $\Delta_{XL}$ falls off as $f^{-2}$, whereas in
the limit of hard cross-links, $gL/f\to\infty$, we find
$\Delta_{XL}=-T/(4fL)$.  In any case, the contribution of the
cross-link is subdominant in the limit of strong pulling force.  Note
that $\Delta_{XL}$ is independent of $\kappa$, which shows us that the
contribution from the cross-link is purely entropic. This contribution
could in fact be evaluated from a directed polymer model which does
not involve any bending rigidity. 

Collecting the leading terms for hard cross-links and strong pulling
force, yields
\begin{equation}
\label{ext}
\frac{\langle z_1 \rangle}{L}-1\;=\;-\frac{T}{(4\kappa f)^{1/2}}
-\frac{T}{4fL}-\frac{D^2}{8L^2}.
\end{equation}
Here we have restored the term due to a finite distance, $D$, between
the chains, which gives rise to a geometric reduction in length due to the
cross-link. This term is presumably unimportant in a network, where
$D/L$ is expected to be small. This term will be neglected in the
following.

\begin{figure}
\includegraphics[width=0.45\textwidth]{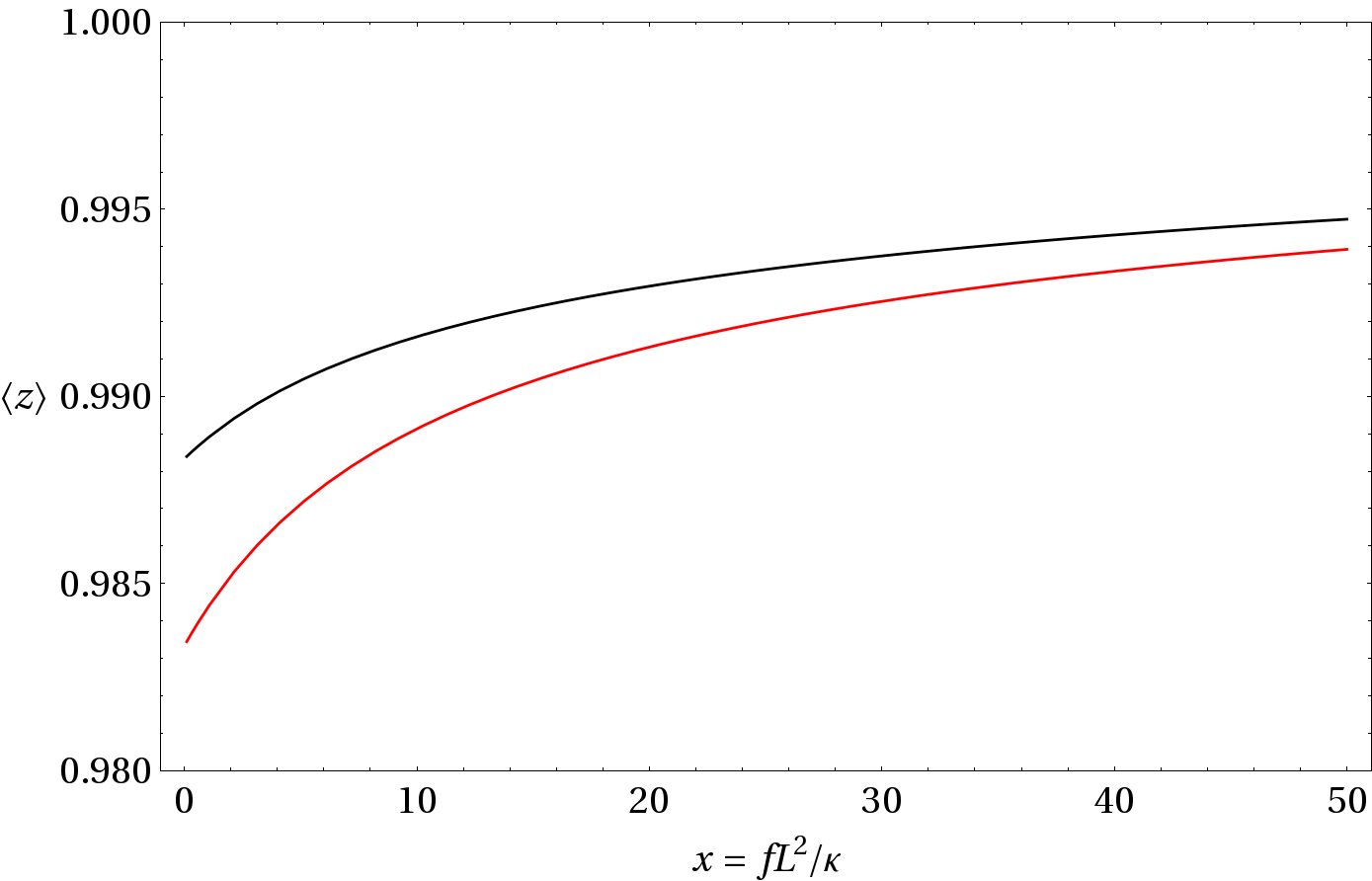} 
\caption{Force-extension curve for $L/L_p=1$; red (black) curve
  without (with) cross-link}\label{fig:force_extension}
\end{figure}

\begin{figure}
\includegraphics[width=0.45\textwidth]{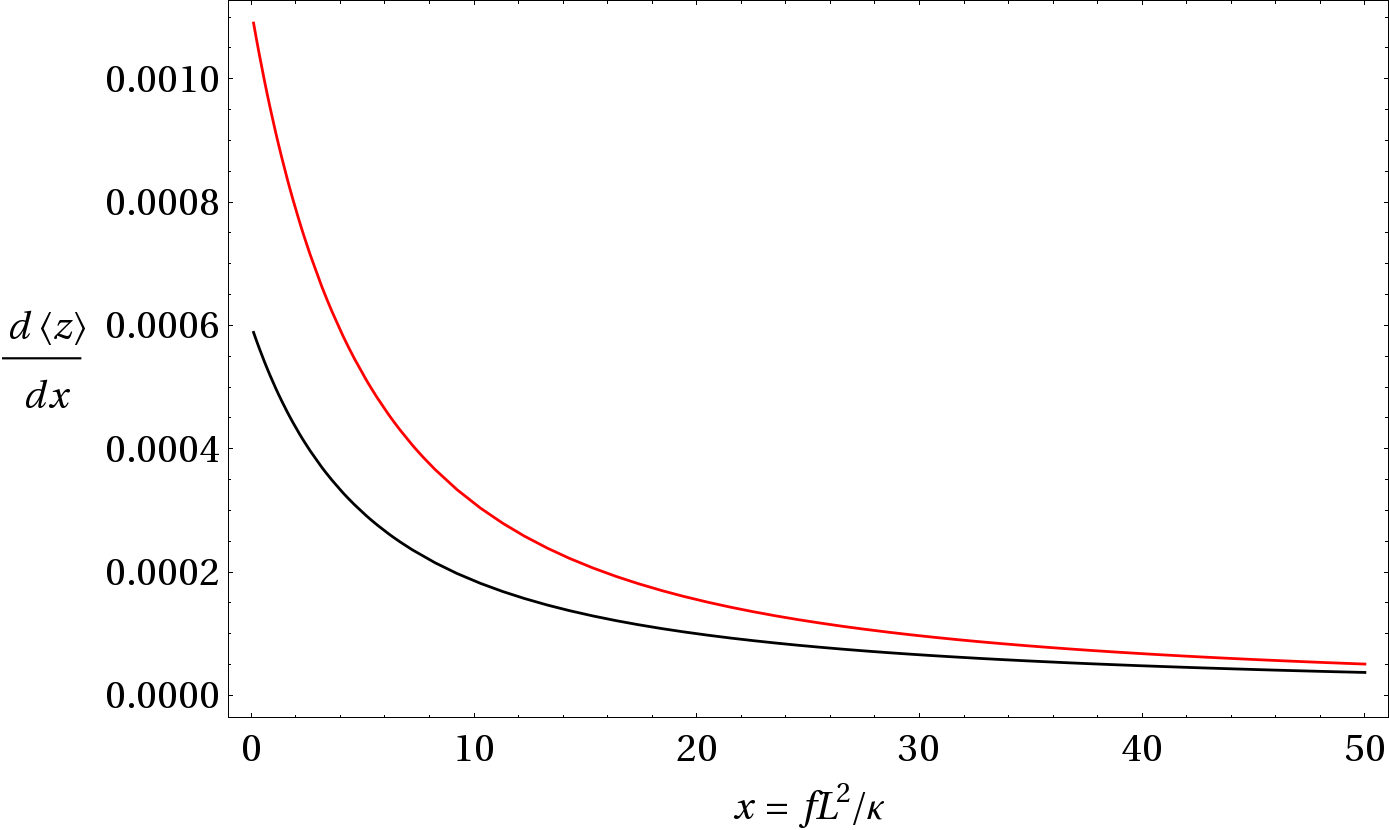} 
\caption{Inverse differential stiffnes for $L/L_p=1$; red (black) curve
  without (with) cross-link}\label{fig:stiffnes}
\end{figure}

The weakly bending approximation is satisfied as long as the r.h.s. of
Eq.~(\ref{full}) is small. We point out that, in the case of $L_p\gg
L$, this approximation is fulfilled even without having very large $x$
(strong stretching). We show the relative extension as a function of
pulling force in Fig.~\ref{fig:force_extension} for hard
cross-links. As discussed, cross-linking enhances the extension due to
the reduction of thermal fluctuations, but the effect becomes less and
less pronounced in the strong stretching limit. It is also of interest
to consider a situation, where a strong pulling force has been applied
and subsequently the change in extension in response to a small change
in the pulling force is measured. This response is determined by the
differential stiffness ${df}/d\langle z_1 \rangle$.  In
Fig.~\ref{fig:stiffnes} we show the inverse stiffness for the same
set of parameters as in Fig.~\ref{fig:force_extension}

As expected the cross-link enhances the differential stiffness. If the chains are
already strongly stretched, the cross-link has little effect. However
for weakly stretched chains the enhancement is considerable.

\section{Randomly cross-linked network}

In this Section, we are going to compute the force-extension relation of a
randomly cross-linked ensemble of oriented chains approximately. We
are guided by the result for the cross-linked pair of chains, which
consists of a single chain contribution, $\sim T/(\kappa
  f)^{1/2}$, and a contribution due to the cross-link, $\sim T/(fL)$, which
is independent of the bending rigidity $\kappa$. We decompose the calculation for the
network accordingly: In the uncross-linked network, the
force-extension relation is determined by the single chain
contribution. To assess the effect of cross-linking, we compute the
free energy difference of the cross-linked network relative to the
uncross-linked system.

Our starting point is the Hamiltonian
\begin{equation}
{\cal H}=\sum_{i=1}^N{\cal H}_0[{\bf r}_i (s) ]
+{\cal H}_\ev+{\cal H}_\Xlink(\mathcal{C}_M)
\end{equation}
with ${\cal H}_0$ given in Eq.~(\ref{WBCH}). To simplify the notation, we
have dropped the subscript ${\bf r}^i_{\perp} (s) \to {\bf r}_i(s) $, such
that ${\bf r}_i$ denotes the transverse excursion of chain $i$. The
excluded volume interaction 

\begin{equation}
{\cal H}_\ev=
\sum_{i<j}\frac{\lambda}{2}\int_0^L ds  \, 
\delta\big({\bf r}_i(s)-{\bf r}_j(s)\big)
\end{equation}
is introduced to balance the attractive interactions due to
cross-linking. The latter are modelled by harmonic springs
\begin{equation}
{\cal H}_\Xlink(\mathcal{C}_M) = 
\frac{g}{2}\sum_{e=1}^M\big({\bf r}_{i_e}(s_e)
-{\bf r}_{j_e}(s_e)\big)^2.
\end{equation}
where
$\mathcal{C}_M := \{i_e,j_e;s_e\}$ is a quenched configuration of $M$ cross-links connecting
polymers $i_e,j_e$ at arclength $s_e$. 

The partition function of the cross-linked system relative to the uncross-linked
melt for a specific realization of
cross-links, $\mathcal{C}_M$, reads 
\begin{eqnarray}
\label{Z_C_M}
Z(\mathcal{C}_M)=\left\langle\exp\!\left(-\frac{{\cal
        H}_\Xlink(\mathcal{C}_M)}{T}\right)\right\rangle. \label{eq:Z(C_M)}
\end{eqnarray}
Here $\langle...\rangle$ denotes the thermal average over all polymer
configurations with the Boltzmann weight $\exp(-{\cal H'}/T)$
of the uncross-linked melt ${\cal H'}=\sum_{i=1}^N{\cal H}_0[{\bf r}_i (s) ]
+{\cal H}_\ev$.

 Physical
observables can be calculated from the quenched-disorder
averaged free energy, $\Delta F=-T [\ln Z]$, where $[...]$ denotes
average over all realizations of random cross-links. We assume that the
number of cross-links can vary and a realization with $M$
cross-links follows the Deam-Edwards distribution \cite{DE}:
\begin{eqnarray}
\label{DE}
P(\mathcal{C}_M)\propto\frac{1}{M!}\left(\frac{{\mu}^2 A}{2N(2{\pi}a^2)}\right)^M Z(\mathcal{C}_M)\;,
\end{eqnarray}
where $a^2\equiv T/g$. The parameter ${\mu}^2$ controls the average number of cross-links per
polymer and the physical meaning of this distribution is that polymer
segments close to each other in the melt have a high probabiblity to
be linked. Of particular interest is the derivative of the free energy
change due to cross-linking with respect to the pulling force 
\begin{equation}
\frac{\partial \Delta F}{\partial f}=-\frac{1}{2}\sum_i\int_0^L ds
[\langle (\partial_s{\bf r}_i)^2 \rangle]
\end{equation}
which yields the mean extension per chain $\frac{1}{N}\sum_i
[\langle z_i(L)-z_i(0) \rangle ]$ relative to the uncrosslinked melt.

The above model is expected to have a gelation transition \cite{UZB}
at a critical cross-link concentration $\mu^2\sim 1$. As far as the
force-extension relation is concerned, we expect to find the single
polymer contribution (Marko-Siggia) below the gelation transition and
a correction due to cross-links above it. The latter will be computed
from the directed polymer model ($\kappa=0$), because it is dominated
by the long wavelength transverse excursions of the polymers which are
correctly captured by the second term in Eq.~(\ref{WBCH}). A similar
mechanism underlies the well-known observation \cite{deGennes} that
the tranverse fluctuations of a strongly stretched wormlike chain are
independent of $\kappa$.






We compute the free energy difference, $\Delta F$, for a network of
cross-linked directed polymers, confined between two planes with their
endpoints free to slide on them. This calculation is analogous to our
previous work on directed polymers \cite{UZB} and some details are
layed out in the Appendix. We point out that our calculation is
restricted to the vicinity of the gel point. Adding the single chain
contribution and denoting the distance from the gelation point by
$\epsilon=\mu^2-1$, we find for the total mean force-extension

\begin{eqnarray}
& &\frac{1}{NL}\sum_i[\langle z_i(L)-z_i(0) \rangle]-1
=-\Delta_{MS}-\Delta_{XL}\nonumber\\
& &\Delta_{MS}=\;-\frac{T}{(4\kappa f)^{1/2}}+\frac{T}{(2fL)}\nonumber\\
& &\Delta_{XL}=\frac{\epsilon^3}{3}\frac{T}{fL+3f^2/g}.
\end{eqnarray}

The above result is quite remarkable in several respects. Whereas the
contribution due to bending shows the $1/\sqrt{f}$ behaviour typical
for WLC, the cross-link contribution is proportional to $1/f$ (hard
cross-links), which is characteristic of freely jointed chains. For
strong stretching, $x\gg 1$, the cross-link contribution has
qualitatively the same dependence on the pulling force, $f$, and the
cross-link strength, $g$, as the corresponding contribution in the
two-chain toy model of the previous Section. There is a singular
contribution to the stretching stiffness at the gelation transition,
which has the same scaling as the shear modulus, namely $\sim
\epsilon^3$. In analogy to the behaviour of the shear modulus in the
well cross-linked regime \cite{UMGZ}, we expect the contribution of the
cross-links to the force extension relation to scale as the density of
cross-links, i.e. to be of the form $\Delta_{XL}\sim\mu^2T/(fL)$.
The excluded-volume interaction, which is included in the gelation theory in
order to prevent collapse of the system upon cross-linking, does not affect
the force-extension relation in the strong stretching regime. Replacing the
excluded-volume interaction of each chain with its neighbours by an
effective harmonic ``cage'' \cite{Nelson}, one notices that for strong pulling
forces the effect of the ``cage'' becomes negligible \cite{Wang}.

\section{Conclusions-Outlook}

In conclusion, we have calculated the effect of random cross-links on
the force-extension relation of an array of parallel-aligned WLCs. Our
calculation is restricted to the strong stretching regime, close to
the gelation transition. The main result is a contribution to the
nonlinear force-extension relation which scales as $\sim 1/f$
for hard cross-links and suppresses the thermal flucuations thus
stiffening the (thermal) stretching modulus. Hard cross-links are
sufficient in the large $L$ limit as the cross-link size $a$ enters
through $TL/(a^2 f)$. Our result is based on a replica field theory of
randomly cross-linked directed polymers originally developed in
Ref. \cite{UZB}. Remarkably, apart from numerical prefactors, the
effect of cross-links is the same as in the case of a simple
two-dimensional model with two chains and a single cross-link in the middle.

An interesting extension of this work would consider cross-links which
are non-local in the $z$-direction. That would imply discontinuities
in the tension of the involved chains as discussed in Ref.\ \cite{MCM}. This problem
will be addressed in a future publication.

\begin{acknowledgments}
We thank the DFG for financial support through SFB 937. PB acknowledges
support by EPSRC via the University of
Cambridge TCM Programme Grant.

\end{acknowledgments} 

\section{appendix}
Here, we outline the calculation of the free energy $\Delta F = - T \, [\ln Z(\mathcal{C}_M)]$, Eqs.~(\ref{eq:Z(C_M)}) and (\ref{DE}) of a disorder averaged randomly cross-linked network relative to the uncrosslinked state. For that, we start with the replica trick:
\begin{equation}
 -\frac{\Delta F}{T} = 
 [\ln Z] = \lim_{n \to 0} \frac{[Z^n] - 1}{n} \label{eq:[lnZ]}
\end{equation}

As usual \cite{UZB} this leads to replicated, $D(n\!+\!1)$-dimensional vectors denoted by a hat, e.\,g.\ $\hat{r} = ( \vec{r}_{0}, ..., \vec{r}_{n} )$. With that, we can express
\begin{align}
 [Z^n] &= \frac{\mathcal{Z}_{n+1}}{\mathcal{Z}_{1}} \quad \text{with}  \label{eq:[Z^n]} \\
 \mathcal{Z}_{n+1} &= \left\langle 
\exp\bigg( \frac{\mu^2}{2NL\phi} \int_0^L ds \sum_{i,j} \Delta(\hat{r}_i(z) - \hat{r}_j(z)) \bigg)
\right\rangle^{\mathcal{H}_0 + \mathcal{H}_\ev}_{n+1} 
\end{align}
In the effective replica partition function $\mathcal{Z}_{n+1}$, the averaging $\langle ... \rangle^{\mathcal{H}_0 + \mathcal{H}_\ev}_{n+1}$ is done with the statistical weights $\mathcal{H}_0 + \mathcal{H}_\ev$, and convenviently, in this form, the disorder average has not to be taken into account anymore. The degrees of freedom are the \emph{replicated} particle positions $\{\hat{r}_i(z)\}_{i=1,...,N}$ and the definition $\Delta(\hat{x}) := \exp(-\hat x^2 / (2a^2) )$ resembles the interaction of the cross-linkers.

In this form the Hubbard-Stratonovich transformation can be used to change the degrees of freedom from the particle positions to a (replicated and Fourier-space) density field $\Omega(\hat q,s) = \langle \frac{1}{N} \sum_i \e^{i \hat q \hat r_i(s)} \rangle$:
\begin{align}
  \mathcal{Z}_{n+1} = \int \mathcal{D} \Omega \; \e^{ -N f_{n+1}(\Omega)} \label{eq:Zn+1}
\end{align}
Here, the effective replica Hamiltonian $f_{n+1}(\Omega)$ is given by:
\begin{align}
 f_{n+1}(\Omega) = f_0 + \frac{\phi^n \mu^2}{2L} \sum_{\hat q \in \text{HRS}} \Delta(\hat q) \int_0^L \dd s |\Omega(\hat q, s)|^2 - \ln \mathfrak{z}
 \label{eq:def:f_n+1}
\end{align}
with the Fourier transform $\Delta(\hat q)$ of $\Delta(\hat x)$ and a single polymer partition function:
\begin{align}
 \mathfrak{z} = \int \mathcal{D}\hat r(s) \exp\!\left( \frac{\phi^n \mu^2}{L} \sum_{\hat q \in \text{HRS}} \Delta(\hat q) \int_0^L \dd s \, \Omega(\hat q, s) \e^{i \hat q \hat r(s) }\right) \label{eq:def:z1}
\end{align}
$\HRS$ stands for the higher replica sector, the set of $\hat q$-vectors with \emph{at least two} non-zero replica components. As been done several times before \cite{GelationReview,UZB}, the excluded volume interaction is assumed to be strong enough to make the network incompressible. Hence density fluctuations, which would be represented by $\Omega(\hat q,s)$ with $\hat q$ having a non-zero component in \emph{only one} replica, do not appear in the Hamiltonian $f_{n+1}(\Omega)$. $f_0$ in (\ref{eq:def:f_n+1}) is an unimportant contribution, which does not depend on $\Omega$.

As next step, we perform the saddle point approximation of (\ref{eq:Zn+1}). The saddle point value of $\Omega$ is given by:
\begin{align}
 \bar\Omega(\hat q, s) = Q \delta_{\vec q_{0} + ... + \vec q_{n}, \vec 0} \int \dd \xi^2 \mathcal{P}(\xi^2,s) \exp\!\left( \frac{\hat q^2 \xi^2}{2} \right) \;.
\end{align}
Here the gel fraction $Q$ is the fraction of chains which are localized, which means they cannot traverse the whole sample, but perform fluctuations around a preferred position. The localization lengths $\xi$ quantify the extent of these in-plane fluctuations, which can depend on the height $s$ in the sample. Their probability distribution $\mathcal{P}(\xi^2,s)$ has been determined in \cite{UZB}. 

We now plug the saddle point value $\bar\Omega$ into Eq.~(\ref{eq:Zn+1}) and restrict ourselves to the vincinity of the gelation transition, i.\,e.\ small gel fractions $Q$. Bearing in mind that $\bar\Omega \propto Q$, we can expand Eq.~(\ref{eq:def:z1}) in powers of $Q$ and easily perform the functional integral over $\hat r(s)$:
\begin{align}
 \mathfrak{z} = 1 + \mu^2 Q + \frac{\mu^4 Q^2}{2!} \mathfrak{z}_{(2)} + \frac{\mu^6 Q^3}{3!} \mathfrak{z}_{(3)} + \mathcal O(Q^4)\;.
\end{align}
Here the coefficients $\mathfrak{z}_{(2)}$ and $\mathfrak{z}_{(3)}$ are given by:
\begin{align}
 \mathfrak{z}_{(2)} = & \sum_{\hat q \in \text{HRS}} | \Delta(\hat q) |^2  
    \int_0^L \frac{\dd s_1 \dd s_2}{L^2} \frac{\bar\Omega(\hat q,s_1) \bar\Omega(-\hat q,s_2)}{Q^2} \nonumber \\
    &\quad \times \exp\!\left(- \frac{\hat q^2}{2f} |s_2 - s_1| \right)  \\
 \mathfrak{z}_{(3)} = & \sum_{\hat q_1,\hat q_2,\hat q_3 \in \text{HRS}} \delta_{\hat q_1+\hat q_2+\hat q_3,\hat0} \Delta(\hat q_1)\Delta(\hat q_2)\Delta(\hat q_3) \nonumber \\
    &\hspace{-20pt} \times \int_0^L \frac{\dd s_1 \dd s_2 \dd s_3}{L^3} \frac{\bar\Omega(\hat q_1,s_1) \bar\Omega(\hat q_2,s_2) \bar\Omega(\hat q_3,s_3)}{Q^3} \nonumber \\
    &\hspace{-20pt} \times \exp\!\left(- \frac{1}{2f} \left(\hat q_1 \hat q_2 |s_1\!\!-\!\!s_2| + \hat q_1 \hat q_3 |s_1\!\!-\!\!s_3| + \hat q_2 \hat q_3 |s_2\!\!-\!\!s_3| \right)  \right)
\end{align}
For simplification, we present the calculation for \emph{hard cross-links}, i.\;e.\ $a=0$. However, the extension to arbitrary $a$ is straightforward. Also, while we present the calculation for a three-dimensional system, the generalization to arbitrary dimension is possible. The result for these generalizations is shown at the end. 

We now perform the sums over $\hat q$. Assuming that the surface area $A$ of the sample in the in-plane directions is large compared to microscopic details of the network, these sums can be changed to integrals. We obtain up to linear order in $n$:
\begin{align}
 \mathfrak{z}_{(2)} = & 1 - n \int \dd(\xi_1^2, s_1) \dd(\xi_2^2, s_2)  \nonumber \\ 
& \Bigg( 
     \ln\!\left( \frac{\xi_1^2 + \xi_2^2 + |s_1 - s_2|/f }{A} \right) \label{eq:res:z2}
    + c  \Bigg) \\
 \mathfrak{z}_{(3)} = & 1 - n \int \dd(\xi_1^2, s_1) \dd(\xi_2^2, s_2) \dd(\xi_3^2, s_3) \nonumber \\ 
& \Bigg( 
     \ln\!\left( \frac{ \xi_1^2 \xi_2^2 + \xi_1^2 \xi_3^2 + \xi_2^2 \xi_3^2}{A^2} \right) 
    + 2c
  \Bigg)  ( 1 + \mathcal O(\varepsilon) ) \label{eq:res:z3}
\end{align}
Here, $\varepsilon := \mu^2 - 1$ is the distance from the sol-gel transition. It is related to the gel fraction by $Q = 2\varepsilon + \mathcal O(\varepsilon^2)$. 
For a better readability, we defined $\int \dd(\xi_\alpha^2, s_\alpha) := \frac{1}{L}\int_0^L \dd s_\alpha \int \dd \xi_\alpha^2 \mathcal{P}(\xi_\alpha^2,s_\alpha)$, and $c = 1 + \ln(2\pi)$ is a numerical constant. 

In a similar fashion, the sum over $\hat q$ can be performed in the second term of Eq.~(\ref{eq:def:f_n+1}).
 \begin{align}
  \frac{1}{Q^2}  \sum_{\hat q \in \text{HRS}} & \Delta(\hat q) \int_0^L \dd s {|\Omega(\hat q, s)|^2} = 1 - n \int_0^L \frac{\dd s}{L} \nonumber \\
 & \int \dd \xi_1^2 \mathcal{P}(\xi_1^2,s) \dd \xi_2^2 \mathcal{P}(\xi_2^2,s) 
  \Bigg( 
     \ln\!\left( \frac{\xi_1^2 + \xi_2^2 }{A} \right)
    + c  \Bigg) \label{eq:res:f}
 \end{align}
As one can see from Eq.~(\ref{eq:[lnZ]}), (\ref{eq:[Z^n]}) and (\ref{eq:Zn+1}), the disorder averaged free energy density is -- in saddle point approximation -- the term of $f_{n+1}(\bar \Omega)$ linear in $n$:
\begin{align}
 \frac{1}{N} \frac{\Delta F}{T} = \left. \frac{\partial f_{n+1}}{\partial n} \right|_{n=0} 
\end{align}
Hence, using the results (\ref{eq:res:z2})-(\ref{eq:res:f}), we can now recompose the free energy up to third order in $\varepsilon$. With spatial dimension $D+1$ and cross-link length $a$, the we obtain:
\begin{align}
 \frac{1}{D N T} \frac{\partial \Delta F}{\partial f} = -  \frac{\varepsilon^3}{3} \frac{L}{Lf + 3f^2 a^2} + \mathcal O(\varepsilon^4) \; .
\end{align}

\end{document}